\documentstyle[preprint,aps]{revtex}
\begin{document}

\LARGE

\begin{center}
{  Universal property of single topological singularity dynamics  }
\end{center}

\normalsize

\begin{center}
{P. Ao\footnote{Corresponding author. E-mail: ao@tp.umu.se }, \\ 
         Department of Theoretical Physics, 
           Ume\aa{\ }University, 901 87, Ume\aa, Sweden },
\end{center}

\begin{center}
{X.-M. Zhu, \\ Department of Experimental Physics, 
           Ume\aa{\ }University, 901 87, Ume\aa, Sweden}
\end{center}

\begin{abstract}
Using the random matrix theory
we demonstrate explicitly 
the insensitivity of the transverse force on a moving vortex 
to non-intrinsic and non-magnetic impurities in a superconductor.
\end{abstract}

%\begin{keyword}
% Write here 3 or 4 keywords separated by semicolons.
vortex dynamics, transverse force, Berry phase
%\end{keyword}

%\end{frontmatter}

% The main text begins here. The \section commands are optional.
\section{Motivation}

The Berry phase acquired by a moving vortex
suggests that the total transverse force 
on a moving vortex is proportional to the superfluid density 
at zero temperature,
irrespective  whether or not impurities are presented \cite{at,tan}.   
This prediction has been confirmed by a 
direct measurement of the transverse
force in dirty superconductors \cite{zhu}.
Though the relaxation time approximation which accounts for impurity effects
in vortex dynamics has been showed to inapplicable \cite{ao},
the drastic reduction the total transverse force by impurities 
were believed right \cite{kv}.

To examine this question further, an exact formulation has been developed 
\cite{az,az2}.
The distribution of impurities is assumed to 
be homogeneous and the variation of impurity potential is only
appreciable on a scale much smaller than size of the vortex core.
If the vortex is allowed to move slowly against the ionic lattice 
background, 
to the leading order of the vortex velocity ${\bf v}_v$, 
the total force acting on  the vortex has the form
$
   {\bf F} = B {\bf v}_v \times \hat{\bf z} -  
            \eta {\bf v}_v  \; .
$
Within the BCS theory, 
the transverse coefficient $B$ of the total transverse force is
explicitly expressed 
in terms of the quasiparticle wave functions:
\begin{eqnarray}
   B &=& i\hbar \sum_{k} \int d^3 x 
     \left\{ f_k \nabla_v u_k(x)\times \nabla_v u^{\ast}_k(x)  \right.
           \nonumber \\
    &&   \left.  -  (1-f_k) \nabla_v v_k(x) \times \nabla_v v_k^{\ast}(x)  
        \right\}\cdot \hat{\bf z} \; . 
\end{eqnarray}
Here  $f_k = 1/(1 + e^{\beta E_k} ) $ is the Fermi distribution function.
The wave functions $\{ \Psi_k(x) \}$ and the corresponding
eigenvalues $ \{ E_k \} $  are determined by 
the usual Bogoliubov-de Gennes equation,
$
   \begin{array}{l}
   {\bf H} \; \Psi_k(x) = E_k \; \Psi_k(x) \; ,
   \end{array}
$
with 
$
   \Psi_k(x) = \left( \begin{array}{c} u_k(x) \\ v_k(x) \end{array} 
       \right) \; 
$   
and the system Hamiltonian  
$
  {\bf H} = \left(\begin{array}{cc} 
                     H  & \Delta  \\
                    \Delta^{\ast} & - H^{\ast} 
                   \end{array} \right) \; .
$
Here $H =  - (\hbar^{2}/2m ) \nabla^{2} -  \mu_F + V(x) $, and
$V(x)$ the impurity potential.
The order parameter $\Delta$ is determined self-consistently.
Now we are ready to explicitly demonstrate that 
the coefficient $B$ for the total transverse force 
is insensitive to impurities as
first observed through the Berry phase calculation.

\section{Demonstration}

In the presence of impurities, the replacement
$\nabla_v \rightarrow -\nabla$ cannot be directly used in Eq.(1)
because of the implicit impurity dependence.
Instead, we expand $\Psi_k$ in terms of eigenfunction
 $\{ \bar{\Psi}_l \} $  of $\bar{\bf H}$, 
a corresponding  Hamiltonian
to ${\bf H}$  without the impurity potential $V(x) $ and with impurity 
averaged $\Delta$, 
\begin{equation}
    \Psi_{k} = \sum_{ l } a_{k l}e^{i\delta_{k l}} \; \bar{\Psi}_{l} \, .
\end{equation}
Here
$ \{ a_{k\mu} \} $ and   $ \{ \delta_{k\mu} \} $ 
are the modulus and phases of the expansion coefficients.
They are functions of the vortex position 
as well as the positions of impurities.
We assume these coefficients to be 
described separately by two independent random matrices,
making use of the randomness of impurities.\cite{random}
$\bar{\bf H}$ still has a dependence on impurities through  the 
impurity averaged order parameter $\Delta$:  
%\begin{equation}
$   \bar{{\bf H} } = \left(\begin{array}{cc} 
                     \bar{ H } & \Delta  \\
                    \Delta^{\ast} & - \bar{H} ^{\ast} 
                   \end{array} \right) $,
%\end{equation}
with $\bar{H} =  - (\hbar^{2}/2m ) \nabla^{2} -  \mu_F $,
$\bar{\bf H} \bar{\Psi}_l = \bar{E}_l \bar{\Psi}_l $.
%and 
%$ \bar{\Psi}_l = \left( \begin{array}{l} \bar{u}_l \\ \bar{v}_l 
%                      \end{array} \right) $ . 
Since $ \{ \bar{\Psi}_\mu \} $ form a complete set, the expansion coefficients
$ \{a_{k l} e^{i\delta_{k l} }  \}$ 
form a unitary matrix, and $\sum_{l \, or \, k} a_{kl}^2 = 1 $.
For $\bar{\Psi}_l $ we can use the replacement $\nabla_v \rightarrow - \nabla$.
Away from the vortex core the value of the 
order parameter is the same  as that in the clean case, guaranteed by
the Anderson theorem.  
Using Eq.(2), 
Eq.(1) becomes
\begin{eqnarray}
   & & B  =  - i\hbar 
      < \sum_k \sum_{l,l'} \int d^3 x  
           \left\{ f_k 
       \nabla_v \left( a_{k l} e^{i \delta_{k l} } \bar{u}_{l}(x) 
         \right) 
        \times   \nonumber \right.   \\
   & & 
       \nabla_v \left( a_{k l'} e^{ - i \delta_{k l'} } 
         \bar{u}_{l'}^{\ast} (x)  \right)  
             - ( 1 - f_k )
       \nabla_v \left( a_{k l} e^{i \delta_{k l} } \bar{v}_{l}(x) 
         \right) \times      \nonumber \\
   & &  \left.  
       \nabla_v \left( a_{k l'} e^{ - i \delta_{k l'} } 
         \bar{v}_{l'}^{\ast} (x)  \right) \right\} \cdot \hat{\bf z} > \; .
\end{eqnarray}
Here $< ... > $ stands for the impurity average 
over the expansion coefficients. 
%The average of impurities
%is carried out by averaging over the ensemble of random matrices.
%After the impurity average, 
%we have 
Then,
\begin{eqnarray}
   B & = & - i\hbar  
       \sum_k \sum_{l}  <a_{k l}^2 > \int d^3 x  \left\{ f_k 
        \nabla_v\bar{u}_l(x) 
        \times  
       \nabla_v \bar{u}_{l}^{\ast}(x)  \right. \nonumber \\
   & &   \left.   -   (1-f_k) 
        \nabla_v \bar{v}_l(x) 
        \times  
       \nabla_v \bar{v}_{l}^{\ast}(x)  \right\} \cdot \hat{\bf z} \; .
\end{eqnarray}
Terms containing derivative to vortex position 
in the expansion coefficients have been averaged to zero. 
Now the replacement of $\nabla_v \rightarrow - \nabla $ can be used, and 
turning the area integral into the line integral 
we have, at zero temperature, the desired result
$
 B  = 2 \pi \hbar \rho_0 \; ,
$
because 
\begin{eqnarray}
  & & \sum_{k, E_k > 0 } \sum_{l} < a_{k l}^2 >
       | \bar{v}_l(|x-x_v| \rightarrow \infty )|^2 \nonumber \\  
  & & =  \sum_{k, E_k > 0 } < |v_k(|x-x_v | \rightarrow \infty )|^2 >
         = \rho_0  \; .
\end{eqnarray}
The above second equality is the Anderson theorem. 
Eq.(5) can also be reached from the envelop 
wave function argumentation\cite{degennes}.  
The insensitivity of the transverse force to impurities can also be 
demonstrated from core state transitions \cite{az2}.

Finally, we point out that 
the impurity effect on the transverse force 
were supposed on the scale of vortex core level space \cite{kv}, 
which is extremely small comparing to the Fermi energy.
There is indeed no effect at all to the superfluid density.

This work was financially supported by the Swedish NFR.

\end{document}